\documentclass[12pt,preprint]{aastex}

\newcommand   {\av}    {\mbox{${\rm A_v}$}}

\newcommand   {\arcs}  {\mbox{$^{\prime\prime}$}}

\newcommand   {\kms}   {\mbox{km\,s$^{-1}$}}
\renewcommand {\ga}    {\mbox{\rlap{\hbox{\lower5pt\hbox{$\sim$}}}\hbox{$>$}}}
\renewcommand {\la}    {\mbox{\rlap{\hbox{\lower5pt\hbox{$\sim$}}}\hbox{$<$}}}

\newcommand  {\solar}  {\mbox{$_{\odot}$}}

\begin{document}



\def\kms {\hbox{km{\hskip0.1em}s$^{-1}$}} 
\def\msol{\hbox{$\hbox{M}_\odot$}}
\def\lsol{\hbox{$\hbox{L}_\odot$}}
\def\kms{km s$^{-1}$}
\def\Blos{B$_{\rm los}$}
\def\etal   {{\it et al. }}                     
\def\psec           {$.\negthinspace^{s}$}
\def\pasec          {$.\negthinspace^{\prime\prime}$}
\def\pdeg           {$.\kern-.25em ^{^\circ}$}
\def\degree{\ifmmode{^\circ} \else{$^\circ$}\fi}
\def\ee #1 {\times 10^{#1}}          
\def\ut #1 #2 { \, \textrm{#1}^{#2}} 
\def\u #1 { \, \textrm{#1}}          
\def\nH {n_\mathrm{H}}
\def\ddeg   {\hbox{$.\!\!^\circ$}}              
\def\deg    {$^{\circ}$}                        
\def\le     {$\leq$}                            
\def\sec    {$^{\rm s}$}                        
\def\msol   {\hbox{M$_\odot$}}                  
\def\i      {\hbox{\it I}}                      
\def\v      {\hbox{\it V}}                      
\def\dasec  {\hbox{$.\!\!^{\prime\prime}$}}     
\def\asec   {$^{\prime\prime}$}                 
\def\dasec  {\hbox{$.\!\!^{\prime\prime}$}}     
\def\dsec   {\hbox{$.\!\!^{\rm s}$}}            
\def\min    {$^{\rm m}$}                        
\def\hour   {$^{\rm h}$}                        
\def\amin   {$^{\prime}$}                       
\def\lsol{\, \hbox{$\hbox{L}_\odot$}}
\def\sec    {$^{\rm s}$}                        
\def\etal   {{\it et al. }}                     
\def\xbar   {\hbox{$\overline{\rm x}$}}         
\newcommand{\agestar}{t_{\star}}
\newcommand{\mstar}{M_{\star}}
\newcommand{\rstar}{R_{\star}}
\newcommand{\tstar}{T_{\star}}
\newcommand{\lstar}{L_{\star}}
\newcommand{\mdote}{\dot{M}_{\rm env}}
\newcommand{\rmaxe}{R_{\rm env}^{\rm max}}
\newcommand{\rmine}{R_{\rm env}^{\rm min}}
\newcommand{\mdisk}{M_{\rm disk}}
\newcommand{\mdotdisk}{\dot{M}_{\rm disk}}
\newcommand{\rmaxd}{R_{\rm disk}^{\rm max}}
\newcommand{\rmind}{R_{\rm disk}^{\rm min}}
\newcommand{\rsub}{R_{\rm sub}}
\newcommand{\rhoconst}{\rho_{\rm cavity}}
\newcommand{\thetacav}{\theta_{\rm cavity}}

\newcommand{\um}{$\mu$m}
\newcommand{\cm}{cm$^{-1}$}
\newcommand{\vlsr}{v$_{LSR}$}
\newcommand{\palpha}{P\,$\alpha$}

\shorttitle{}
\shortauthors{}

\title{Massive Star Formation of  the Sgr~A East HII Regions\\
 Near the  Galactic Center}
\author{F. Yusef-Zadeh\altaffilmark{1,2},
J.H. Lacy \altaffilmark{2,3},
M. Wardle\altaffilmark{4},
B. Whitney\altaffilmark{5},\\
H. Bushouse\altaffilmark{6},
D. A. Roberts\altaffilmark{7},
R.G. Arendt\altaffilmark{8}
}
\altaffiltext{1}{Department of Physics and Astronomy, Northwestern University, Evanston, Il. 60208}
\altaffiltext{2}{Visiting Astronomer at the Infrared Telescope Facility,
which is operated by the University of Hawaii under Cooperative Agreement
no. NNX-08AE38A with the National Aeronautics and Space Administration,
Science Mission Directorate, Planetary Astronomy Program.}
\altaffiltext{3}{Department of Astronomy, University of Texas, Austin, TX 78712}
\altaffiltext{4}{Department of Physics and Astronomy, Macquarie University, Sydney NSW 2109, Australia} 
\altaffiltext{5}{Space Science Institute, 4750 Walnut Street, Suite 205, Boulder, CO 80301}
\altaffiltext{6}{STScI, 3700 San Martin Drive, Baltimore, MD  21218}
\altaffiltext{7}{Adler Planetarium and Astronomy Museum, 1300
South Lake Shore Drive, Chicago, IL 60605}
\altaffiltext{8}{University of Maryland -
Baltimore County, GSFC, Code 665, Greenbelt, MD 20771}


\begin{abstract} 

A group of four compact HII regions associated with the well-known 50 \kms\ molecular cloud is the closest site of 
on-going 
star formation to the dynamical center of the Galaxy, at a projected distance of $\sim$6 pc. We present a study of 
ionized gas based on the [NeII] (12.8\,\um ) line, as well as multi-frequency radio continuum, HST Pa$\alpha$ and 
Spitzer IRAC observations of the most compact member of the HII group, Sgr~A East HII  D. The radio continuum image at 
6cm shows that 
this source breaks up into two equally bright ionized features, D1 and D2. The 
SED of the  D source is  consistent 
with it being due to a 25$\pm3$   \msol\, star with a luminosity of 
$8\pm3\times10^4$ \lsol. 
The inferred mass, effective 
temperature of the UV  source and the ionization rate are compatible with a young O9-B0 star. 
The ionized 
features D1 and D2 are considered to be ionized by UV radiation collimated by an accretion disk. We consider that 
the central massive star 
photoevaporates its circumstellar disk on a timescale of 3$\times10^4$ years giving a mass flux $\sim 
3\times10^{-5}$\,M\solar\,yr$^{-1}$ and producing the ionized material in D1 and D2 expanding in
an inhomogeneous medium. 
The ionized gas kinematics, as traced by the [Ne II] emission,
is difficult to interpret, but it could be explained by the   
interaction of a bipolar jet with surrounding gas along with  
what appears to to be a conical wall of lower velocity gas. 
The other HII regions, Sgr~A East A-C, 
have morphologies 
and kinematics that more closely resemble  cometary flows seen in other compact HII regions, where gas moves along a 
paraboloidal surface formed by the interaction of a stellar wind with a molecular cloud.
\end{abstract}

\section{Introduction}
\label{introduction} 

The Galactic center region is known to have a high concentration of
massive, warm, dense and turbulent molecular clouds.
Due to strong tidal forces exerted by the
gravitational potential of the nuclear cluster and the massive black
hole, Sgr~A*, only dense molecular gas is expected to survive in this
region \citep[see][]{morris96}. 
This has the ramification that the formation of massive
stellar clusters must be pervasive in this region as evidenced by the
Arches and Quintuple clusters as well as the nuclear massive cluster
centered on Sgr~A* \citep[][and the references therein]{figer02,bartko09}.
There are also young
clusters of O-B stars traced by ultracompact (UC) HII regions in Sgr B2
\citep[e.g.][]{depree98} which is possibly the best  example of mini-starburst
activity in the Galaxy.  Although our present understanding of star
formation is primarily limited to observations of low-mass stars
\citep[][]{shu87,mckee07,zinnecker07}, probing star forming sites in
the Galactic center region can be useful in addressing the key mechanism
by which massive stars are formed (M $>$ 10 \msol) \citep[see][and references cited therein]{hoare07}.

Here, we examine
the nature of a group of massive young stars lying within  2.5$'$ of
the dynamical center of the Galaxy 
\citep[][]{reid04,ghez08,gillessen06}. 
At a projected distance of 6 pc from Sgr~A*, there is a cluster of compact HII
regions on the edge of the 50 \kms\ molecular cloud M-0.02-0.07.
This cluster consists of  brightest HII regions in  the Galactic center with the exception of 
the cluster of HII regions in Sgr B2.  
The 50 \kms\  molecular cloud itself is interacting with the
nonthermal Sgr~A East \citep[e.g.][and references cited therein]{tsuboi09}  
which is known to be a supernova remnant (SNR G0.0+0.0).
Since Sgr~A East appears to be interacting with the Galactic center circumnuclear molecular ring,
which itself orbits  Sgr~A*, it is probable that Sgr~A East, M-0.02-0.07,
and the cluster of HII regions all lie near the Galactic center,
not far from their projected distances.
The compact HII regions, known as the Sgr~A East HII A to D (Sgr A A-D),
trace the site of recent massive star formation nearest to the dynamical center 
of the Galaxy. The four HII regions have radial velocities ranging
between 43 and 49 \kms, thus appear
dynamically coupled to the 50 \kms\ molecular cloud \citep[][]{goss85,serabyn92}.  
With the exception of [ArIII] emission from Sgr~A-D HII source, 
IR spectroscopic measurements
detected [NeII] and [ArIII] line emission from all four components
(A-D) indicating that the exciting stars have spectral types of O8-B0
\citep{serabyn92},
consistent with H76$\alpha$ measurements \citep{goss85}. 

The measured extinction toward Sgr~A East A-C is less than 
that of source D \citep{cotera99,serabyn92}. 
Additional extinction for
source D is likely due to the 50 \kms\ molecular clouds or 
due to material associated with the ionizing stellar source 
\citep{serabyn92}. 
These authors estimate the extinction values at 12.8$\mu$m 
are  $\sim1-1.3$mag for A-C and  $\sim$3.3 toward the Sgr A D source.  
These extinction measurements imply that the Sgr~A 
East A-C HII regions, which are resolved spatially, are located on the front
side of the molecular cloud, whereas the compact Sgr~A  D source or
G359.96-0.08 (hereafter this source is called  D), may be
embedded within the 50 km s$^{-1}$  molecular cloud.  Near-IR measurements
have also examined the stellar sources associated with 
these HII regions \cite{cotera99}. A highly  reddened source with a
H-K$^\prime$ $\sim$ 3.5 is detected at the position of the most compact
HII source D.  The low fractional abundance of Ar$^{++}$ led to the
suggestion that stellar temperatures of the Sgr~A East HII regions
range between 34,500 and 36,500K \citep{serabyn92} which are
consistent with spectra types O8-O9 stars \citep{goss85,zadeh89}. The spectral identification of
the Sgr~A D source has not been clear but \citet{cotera99} suggest
that its spectrum could be consistent with a combination of an
ultracompact (UC) HII region and B[e] stars. Here, we investigate
radio and infrared properties of the most compact (D source)
and extended (A source) of the Sgr~A East HII regions. 

\section{Observations}
\subsection{Radio Continuum}

Radio continuum observations were made with the D configuration of
the Very Large Array (VLA) of the National Radio Astronomy
Observatory\footnote{The National
Radio Astronomy Observatory is a facility of the National
Science Foundation, operated under a cooperative agreement by
Associated Universities, Inc.} (NRAO), at 43 GHz (7mm).  
The source was observed as part of a survey
of UC HII regions in the Galactic Center, which took place May 29,
2009.  The source was observed in the fast switching mode of the VLA
(each cycle had 150 seconds on the source and 15 seconds on the
complex gain calibrator - NRAO530) for a total of 20 minutes.  3C286
was used as a flux calibrator.  Standard data editing and calibration
of the gains was done in AIPS. Using standard calibration,  we also 
reduced multi-configuration radio continuum data at 8.3 GHz (AY43). 
The 4.8GHz data were taken from \citet{zadeh93}.

\subsection{[NeII] Line \& 12.8$\mu$m Continuum}
The Sgr~A East HII region cluster was observed in the [Ne II] (12.8 \um )
line with the high-resolution mid-infrared spectrograph TEXES
on the NASA IRTF on 2009 June 3 and 2010 May 30 (UT).
TEXES \citep{lacy02} has several operating modes.
These observations were made in two modes: high-resolution,
cross-dispersed, with velocity resolution $\sim$4 \kms\ and
spatial resolution $\sim$1.2\arcsec\ 
along a 1.4$\arcsec\times$7.5\arcsec\ slit, and medium resolution,
long-slit, with velocity resolution $\sim$25 \kms\ along a
1.4$\arcsec\times$55\arcsec\ slit.
The pixel spacing is 0.36\arcsec\ spatially and 0.95 \kms\ spectrally
in cross-dispersed mode.  It is 0.36\arcsec\ and 8.9 \kms\ in long-slit mode.
Maps were made by stepping the N-S oriented slit to the east,
with 0.7\arcsec\ steps in 2009 and 0.35\arcsec\ steps in 2010.
The Sgr~A East HII region was mapped in 2009 in cross-dispersed
mode by making a series of scans, each covering
30$\arcsec\times$7.5\arcsec\ (RA$\times$Dec), which were pieced together
to cover a 34$\arcsec\times$68\arcsec\ region (with some small gaps),
with the help of a long-slit scan that covered the entire region.
HII source D was reobserved in 2010 to obtain better spatial sampling
and higher signal-to-noise ratio.
These  measurements also allowed us to detect  12.8$\mu$m emission from the D source using 
our low  spectral resolution data and medium resolution data taken in 2009.  

The spectral coverage of the data was 2270 \kms , which in the 
cross-dispersed observations was broken into 8 spectral orders of
the high-resolution grating.
[Ne II] line emission was detected in a 60 \kms\ interval centered
at \vlsr\ $\approx$ 46 \kms .
Dust continuum emission was only marginally detected outside of this
spectral interval (Also see Figure 3 showing source D before and after deconvolution.)

The data were first processed with the standard TEXES pipeline reduction
program \citep{lacy02}, which removes spikes, corrects for optical
distortions, and produces wavelength and intensity calibrated maps.
It was then processed with a maximum-entropy-method deconvolution
routine, which improved the spatial resolution to about 0.7\arcsec ,
slightly better than the Infrared Telescope Facility (IRTF) diffraction limit.
Because a part of the point-spread function is caused by seeing,
which cannot be exactly the same for the observations of the PSF
calibrator (VX Sgr) and the HII region, the results of the deconvolution
should be viewed with some caution, but the close similarity between
the deconvolved [Ne II] data and the higher resolution radio continuum
data indicates that the results should be reliable.
In addition, all of the features that we discuss in the [Ne II] data
can be seen before deconvolution, although some sources, notably the
two peaks in HII source D, blend together.
The absolute astrometry of the 12.8$\mu$m data is about few arcseconds. 
We obtained coordinates for the [Ne II] maps by aligning them with
the radio maps.  These should be good to better than 0.35" (1 pixel).
The relative astrometry of the [Ne II] and 12.8um continuum should
be better since they were observed simultaneously.

\subsection{1.90$\mu$m Continuum and Pa$\alpha$ Line Data}

Pa$\alpha$ emission-line and continuum images of the Sgr~A East region
were obtained as part of the larger survey of the Galactic center by
\citet{wang10}, using the Near-Infrared Camera and Multi-Object
Spectrometer (NICMOS) on HST (program 11120). Exposures were obtained 
in March--May 2008 using the NICMOS camera 3, which has a pixel scale 
of 0.20\arcsec. 
Images were obtained in narrow-band filters centered on the
Pa$\alpha$ 
emission line at 1.87$\mu$m and the nearby continuum at
1.90$\mu$m. A pattern of dithered observations was obtained at each
pointing in the survey with each exposure having a duration of 48 seconds.

For this study we used a total of 8 exposures in each of the two filters
that covered the region near source D. Calibrated images
were retrieved from the Hubble Data Archive, which had been processed
using the latest NICMOS instrumental calibrations and data reduction
software. The eight calibrated images in each filter were combined using
the "multidrizzle" task in the STSDAS package of PyRAF. The final
drizzle-combined images at Pa$\alpha$ 
and 1.90$\mu$m were subsequently
rescaled using the latest NICMOS photometric zeropoint information to
produce images in calibrated units of Jy.

To determine the position uncertainty  of
the stars,  a Gaussian was fitted to two different  
F190N  images with different exposures. Both images were used in  making the drizzled
mosaic.  This allows us to 
to measure random pointing errors from exposure to exposure, as well as the
uncertainties in our measurements of the centroid in the images.
The random errors in pointing and measurements are on the
order of $\sim0.1''$. This is entirely consistent with the
random relative errors associated with HST reacquiring guide
stars in between exposures.
In addition to this random error of $\sim 0.1''$, there is  also
an absolute rms uncertainty of $\sim0.2''$  in how well the
coordinates of the HST guide stars are known. 
Thus, the coordinates of the HST images can be trusted only at the
level of 0.2\arcsec.

\subsection{IRAC Measurements}

IRAC observed the SGR A D as part of the Galactic center (GALCEN) survey by \citet{stolovy06}. It was covered by 5 exposures with 1.2 sec frame times.
The IRAC flux densities for Sgr A D were obtained from both the GALCEN Catalog
\citep{ramirez08} and the independently reduced GLIMPSE II catalog (an
extension of the GLIMPSE survey, \citet{benjamin03}). At 3.6, 4.5, and 5.8
$\mu$m we use the GLIMPSE II measurements, which tend to be $\sim 15\%$ lower
than the GALCEN measurements. The GALCEN catalog indicates that the source is near
or at saturation at all wavelengths, although at 3.6 $\mu$m the source is only
slightly past the nominal saturation limit. The GLIMPSE II catalog does not
include an 8 $\mu$m flux density. At this wavelength (and only this wavelength) 
the IRAC images show a local minimum at the peak of the source which is
characteristic of a strongly saturated source. We have attempted to apply T.
Jarret's IRACWORKS 
software\footnote{http://ssc.spitzer.caltech.edu/dataanalysistools/tools/contributed/irac/iracworks/}
to recover the flux of Sgr A D from the saturated images. At 3.6 - 5.8 $\mu$m
IRACWORKS does not recognize Sgr A D as a saturated source, suggesting that
cataloged flux densities are not less than $\sim 80\%$ of the true values. At 8
$\mu$m IRACWORKS successfully fits the source for only 2 out of 5 of the
exposures, but does find a consistent result of $5510 \pm 450$ mJy, which is about
2.5 times higher than the lower limit provided by the GALCEN catalog.
Reported magnitudes and flux densities are listed in Table 1.

\begin{deluxetable}{llll}
\tablewidth{0pt}
\tablecaption{2MASS, IRAC and TEXES Flux Densities of  Source D}
\tablehead{
\colhead{Source}&
\colhead{Band}&
\colhead{Magnitude}&
\colhead{Flux Density (mJy)}
}
\startdata
2MASS 17455154-2900231\\
& J		& $>17.607$					& 	$<0.1$\\
& H 	& $>15.085$					& 	$<0.947$\\
& K		& $11.20 \pm 0.038$ 			& 	$22.08 \pm 0.7$\\

SSTGC 0556837 (GALCEN)\tablenotemark{a}\\
& 3.6		& $7.559 \pm 0.009$		& $266.0 $\\
& 4.5		& $5.571 \pm 0.004$		& $1062 $\\
& 5.8		& $4.152 \pm 0.006$		& $2511 $\\
& 8			& $3.617 \pm 0.006	$		& $2292 $\\

SSTGLMA G359.9672-00.0811 (GLIMPSE II)\\
& 3.6		& $7.724 \pm 0.055$		& $228.4 $\\
& 4.5		& $5.755 \pm 0.071$ 		& $896.7 $\\
& 5.8		& $4.346 \pm 0.016$		& $2101 $\\
& 8			& \nodata				& \nodata\\

(IRACWORKS)\\
& 8			& $2.66 \pm 0.09$			& $5510 \pm 450$\\

(TEXES)\\
& 12.8			& 		& 7500\tablenotemark{a} $\pm 1500$
\enddata
\tablenotetext{a}{May be affected by saturation} 
\label{tab:irac_2mass}
\end{deluxetable}

\section{Results}

The large-scale distribution of thermal and nonthermal emission from 
Galactic center is complex.  However, within 4$'$ of Sgr~A*, the
distribution of dust, molecular, ionized, and synchrotron emission
shows that the 50 \kms\ GMC M-0.02-0.07 interfaces with the supernova
remnant Sgr~A East on one side and the cluster of HII regions on the
opposite side.  
Figure 1a,b show the distributions of the 450 and 850\,\um\ 
thermal emission from dust, superimposed on a 20cm  radio
continuum image, respectively \citep{zadeh87,tsuboi09,price00}.
Contours of submm emission 
arises from the eastern edge of the 50 \kms\ molecular cloud. 
These figures show that the cluster of compact
HII sources Sgr~A  A thru D are distributed 
in a region relatively devoid of emission at 450 and 850$\mu$m. 
The cavity in the submm distribution 
might be caused by a drop in dust temperature. 
We made several cuts across the region  where there is a depression 
of submm emission and determined a constant 450$\mu$m to 850$\mu$m  flux ratio. 
Since the emission is optically thin,  this ratio corresponds to a constant temperature.
We note that  dust temperature can not be constrained  because 
it is entangled with the  uncertainty in the ratio of optical depths. 
Since there is no evidence for a temperature gradient in the dust distribution, 
it is possibile that 
the gas in the cavity is converted into formation of stars within the cloud. 
Source A is the most extended HII feature and lies to the north,
whereas source D, the most compact HII source, lies to the south.
The distribution of the four HII regions appear to be anticorrelated
with the distribution of dust and molecular gas, consistent with
kinematic evidence that the 50 \kms\ GMC is the parent cloud from
which a cluster of massive young stars recently formed.

Figure 2a shows a grayscale image of free-free continuum emission from the Sgr~A East HII cluster at 
4.8\,GHz and Figure 2b shows the integrated [NeII] line intensity map. The spatial distribution of [NeII] 
line emission matches very closely the free-free 4.8\,GHz continuum distribution, demonstrating that [Ne 
II] is a good tracer of the ionized gas. Figure 2c shows a contour map of source D before and after 
deconvolution demonstrating the elongated structure of source D in both maps. 
Figure 2d shows the line and 
continuum maps in two different panels. 
The continuum map is from a 1200 \kms  interval blueward of the line. 
The continuum map (right panel of Fig. 2d) is narrower in the E-W direction 
than the line, 
and is centered on the peak of the line map within about 0.5$''$ of  the centroid of the line emission.  
Given the low resolution of the 2009 data, we can not determine the relative position of the continuum and line 
peaks with respect to each other. The high resolution data can not detect 
the continuum emission. 

In this paper, we discuss two of the HII regions in the Sgr A East group:
D, the bright, compact source to the south, and A, the most extended 
source to the north.  Sources B and C have morphologies and
kinematics similar to source A, but are more irregular and fainter.

\subsection{The  D Source}

The brightest radio continuum source
in Figure 2 is source D, which is resolved for the first
time into two equally bright radio continuum sources, D1 (East) and D2 (West),
separated by 1\dasec2 at Position Angle (P.A.) $\approx$ 70\deg .
Each radio source consists of a compact source and a resolved
structure elongated in the north-south (NS) direction. Contours of the
emission at 4.87 GHz, as shown in Figure 3a, extend for $\sim18''$ or
$\sim$0.7 pc, assuming that they are located at the Galactic center
distance of 8 kpc.  The Sgr~A East HII regions have  also been observed
at 8.3 and 43 GHz.  Figure 3b shows a grayscale image of source D at
43 GHz with a spatial resolution of 0\dasec58 $\times$ 0\dasec47.  Both
sources D1 and D2 show N-S structure with an extent of 4\asec - 5\asec.
The elongated features  of sources D1 and D2  each show 
a compact  component at 43 GHz. 
Table~2 give the 
integrated flux, the size and the peak flux of
individual sources D1 and D2 at 4.87, 8.3 and 43.3 GHz,
respectively.   The Gaussian fitted positions of D1 and D2 are 
$\alpha\, ,  \delta\,  (J2000) = 17^h 45^m 51^s.62,  -29^0 00' 22''.59$ 
and  $17^h 45^m 51^s.53,  -29^0 00' 22''.82$, respectively. 
The images from which the fitted data are obtained have
identical spatial resolution 0\dasec8 $\times$ 0\dasec4 at all three
frequencies.  
The integrated  fluxes  S$_{\nu}$ for  source D1 and D2 given 
similar spectral index of $\beta\sim0.1$ (where 
S$_{\nu}\propto\nu^{\beta}$) between 8 and 44 GHz. 


\begin{deluxetable}{lrrrrrrrrrrrrrrrrrrrrrrrrr}
\tablecaption{Parameters of the Fit to Source D G359.96-0.08}
\tablewidth{0pt}
\tablehead{
\colhead{Source} &
\colhead{$\nu$ (GHz)} &
\colhead{S$^a_\pm\sigma$(mJy)} &
\colhead{$\theta_{maj} \arcsec$} & \colhead{$\theta_{min} \arcsec$} &
\colhead{S(mJy)$_{peak}\pm\sigma$} 
}
\startdata
 D1  & 4.87 &    41.4$\pm$1.1  &1.7  & 1.2 & 6.6$\pm0.16$ \\
 "  &  8.45 &  32.3$\pm$2.3    & 1.7 & 0.7 & 8.6$\pm0.5$ \\
 "  &  44.34 & 38.9$\pm$3.6 & 1.5 & 1.0 & 8.7$\pm0.7$\\

\\
 D2  &  4.87 & $46.2\pm$1.1 & 1.9 & 1.2 & 7.6$\pm0.16$ \\
 "  &   8.45 &  27.3$\pm$1.8   &  1.4 & 0.5 & 10.0$\pm$0.5\\
 "  & 44.34 &  29.4$\pm$2.2 & 1.1 & 0.5 & 12.5$\pm$0.7 \\

\\
\enddata
\tablenotetext{a}{The spatial resolution is 0.83\arcs\ and 0.52\arcs\ (PA=11$^\circ$)} 
\end{deluxetable}

From the radio images, D1 and D2 appear to be a pair of stars,
each of which has a north-south bipolar jet.
This situation could occur
if each star has a disk collimating its jet, and the two disks
are coplanar, perhaps as a result of forming  from the same cloud.
But for this to be the right model, the two stars would have
to be nearly identical in their luminosities and jet properties,
including a difference in their N and S jets, as well as having
closely aligned disks.  This seems highly unlikely (see additional  arguments
against two stars, as described below).  A model with
one exciting source for the entire ionized gas distribution would
seem more plausible if it could explain the spatial distribution
of radio emission.

To examine the nature of the exciting star or stars in source D further,
we compared the radio images
with a Pa $\alpha$ and 1.90$\mu$m images of the region. 
The cross drawn on Figure 3a shows the peak position of the 
1.90$\mu$m emission.  The position of the stellar
source  does not change when both 1.90 and 1.87$\mu$m images are compared with each other. 
The centroids of the source at these wavelengths match to $0.02"$.
Since the  images in the these two filters were obtained back-to-back,
without moving the telescope, the relative positioning should be
within 10-20 milli-arcseconds.
The 1.90$\mu$m  emission appears to coincide close to 
the compact
component of source D2, suggesting
that a single source in the vicinity of or  coincident with D2 may be responsible
for the ionization producing the extended radio continuum and
Pa$\alpha$ nebular emission. Alternatively, 
it is possible that the extinction is variable locally 
being responsible for 
a lack of  1.90$\mu$m detection from source D. 
In the continuum image the D2 source is very bright 5.1$\pm$ mJy while D1 is nearly undetected $<0.9\pm0.1$ mJy.
In the emission-line image, on the other hand, D1 and D2  sources have
the flux of 3.7$\sim$0.2 and 7.8$\sim$.2 mJy, respectively.  
If the extinction could be causing the D1 source to
be hidden, then one would expect that both sources 
would be  equally hidden in
both bands, unless the P$\alpha$  flux at the location of
D1 is actually coming from something other than D1. 
If D1 were intrinsically as bright as D2 at 1.90$\mu$m, but
being hidden by extinction, then the ratio of their
fluxes gives an equivalent magnitude difference of
1.9 mag due to extinction.  We can make the same analysis for the 
hypothetical D1 source using IRAC measurements. 
If a point source is present at D1, then its brightness
must be at least 40 times  lower than the source at D2. 
Otherwise, we  would detected  the IRAC 
source looking slightly extended, and the position would be closer to the midpoint between D1 and D2.
That means that a source at D1 must be at least 1.5-2.5 mag fainter than the source at D2.

In order to characterize the properties and the evolutionary phase
of the stellar source responsible for the excitation of ionized
material in source D, we obtained the Spectral Energy Distributions
(SEDs) of the compact source closest to D2 using measurements from 2MASS,
IRAC and Pa$\alpha$ and 1.90$\mu$m  of NICMOS/HST from 1.24$\mu$m to 8$\mu$m.
The background subtracted Gaussian fitted position of  the bright 1.90$\mu$m source 
gives  the following coordinates 
  $\alpha\, ,  \delta\,  (J2000) = 17^h 45^m 51^s.545,  -29^0 00' 23''.420$. 
Similarly, the peak  radio position D2 at 43 GHz is at   
$\alpha\, ,  \delta\,  (J2000) 17^h 45^m 51^s.536,  -29^0 00' 23''.064$. 
The 1.9$\mu$m fitted position gives an $\alpha$, $\delta$ offset of $\sim 0.12''$ and 0.36$''$, respectively, 
southeast of the peak radio emission in D2. 
Given the random pointing errors and absolute uncertainty in the 1.9$\mu$m positions  $0.2''$, 
it is likely  that the  1.90$\mu$m source  could lie  between D1 and D2 but closer to D2 by $\sim0.3$'' 
than to D1. 
Due to the poor resolution of the 12.8$\mu$m continuum data, we can not identify the relative position 
of the continuum emission with respect to sources D1 and D2. 
Lastly, the IRAC GALCEN and GLIMPSE II catalogs identify the position of the IRAC source as
$\alpha\, ,  \delta\,  (J2000) 17^h 45^m 51^s.55,  -29^0 00' 23''.0$. 
with a 0.2$''$ and 0.3" uncertainty, respectively.  The centroid of the IRAC source is shown 
as a triangle with an uncertainty of 0.2$''$ in Figure 3a.
This makes it very likely that the 
IRAC source corresponds to D2, and not D1. D1 and D2 could have not been separately 
resolved by IRAC, but if there were two sources of equal brightness, the point source 
would have appeared  slightly more blurred and more centrally located than observed. 
Given the positional uncertainty, we can't rule out the possibility of a single source
that is centrally located between D1 and D2, but an association with D2 is most  likely.
Here, we  name source D or G359.96-0.08 interchangeably between the 
central star and the peak of the radio emission.

Figure 4 shows  IRAC images of the Sgr A East A-D regions
at 3.6, 4.5, 5.8, and 8 $\micron$ with Sgr A D source being saturated 
at 8 $\micron$ \citep{stolovy06,ramirez08,arendt08}. 
The associated infrared dark cloud is best seen to the west of 
the HII regions where submm emission is most prominent, as seen 
in Figure 1.  
D1 and D2 could have  not been separately resolved by IRAC, but if there 
were two sources of equal brightness, the point source would appear
slightly  more blurred than observed. IRAC data indicate that 
the peak of the  point source does seem to match closer to D2 than D1. 
We used lower limits to the saturated IRAC fluxes in the SED fitting. 
The SED
of compact component of D2 is analyzed by comparing to a set of SEDs
produced by a large grid of YSO models \citep{robitaille07,whitney03}.  A linear regression fitter was used to find
all SEDs from the grid of models that are fit with the assumption that
the difference between the $\chi^2$ per datapoint and the $\chi^2$ per
datapoint of the best fit is less than three.  The physical parameters
corresponding to those models are averaged along with their standard
deviations.  Figure 5 shows the fit to the SED of the  star associated with 
source D or G359.96-0.08, and
Table 3 gives the mean values of visual extinction, stellar mass,
luminosity and temperature based on data from 2MASS, IRAC, 12.8$\mu$m continuum, 
and upper limits at 450 and 
850$\mu$m.  
These physical characteristics are
consistent with the assertion that  the central star of 
G359.96-0.08  is a hot, massive
star.  This star ionizes the surrounding gas, producing thermal radio
continuum with a total integrated flux of 81 and 89 mJy at 43.3 and
4.87 GHz, respectively. From the radio continuum and SED fitting
using infrared continuum emission, 
it appears that the central  star is of  spectral type 
 ranging between $\sim$B0 and  O9 with a  
luminosity V class \citep{vacca96}  
can account for the ionization of the gas in source D, including
both D1 and D2.

\begin{deluxetable}{lrrrrrrrrr}
\tablecaption{The Parameters of the SED Fit to  G359.96-0.08}                  
\tablewidth{0pt}
\tablehead{
  \colhead{Source} &
  \colhead{$< \av >$} &
  \colhead{$< \mstar >$} &
  \colhead{$ < \lstar > $} &
  \colhead{$ < T_* >$}
}
\startdata
D2 &      42.9$\pm5.8$ &   24.8$\pm$3.4 &  8.1$\pm2.6$E+04 & 3.6$\pm0.6$E+04\\
\enddata
\end{deluxetable}


Figure 6 shows four channel maps of [NeII] line emission,
each integrated over a velocity range of 15 \kms.
As is true of the radio continuum distribution, the [Ne II] emission
is symmetric about a north-south line between D1 and D2, but only
if the velocities are reflected about \vlsr\ $\sim$ 47 \kms.
In an alternative display,
the [Ne II] data cube is displayed in Figure 7 as a series of
position-velocity diagrams.
The panels in Figure 7 are cuts in right ascension at six different declinations,
separated by 1.1\arcsec.
In the cut passing at declination of 
$-29^0 00' 23.3''$, 
south of the peak emission, the line emission, as
shown in the   third  panel from left, is
narrow, both spectrally and spatially.  
The fourth panel from left  shows clearly  two blended peaks,
one at 17$^h$45$^m$51.60$^s$, 
and one at 51.56$^s$.
Following the naming used for the radio peaks, we will refer to them
at D1 and D2.
D1 peaks at \vlsr\ = 48 \kms , and has a red shoulder extending to
60 \kms.  D2 peaks at 45 \kms , and has a blue shoulder
extending to 35 \kms.
Going to the north, the spatial and spectral separation between D1 and D2
continues to increase, and the red and blue shoulders become more
prominent.
In the fifth panel from left, about  1.5\arcsec\ north of the peak,
at $-29^0 00' 21.2''$, 
the emission pattern forms an almost complete ring in the position-velocity diagram.
All of the position-velocity diagrams show the same symmetry seen in the channel maps,
being unchanged under a reflection through RA=
17$^h$45$^m$51.58$^s$, and 47 \kms.


The strong symmetry of the ionized gas kinematics about a line
between D1 and D2 makes it even less likely that these peaks could
be ionized by two stars or that they could be ionized by a star
centered on D2.  It is possible that the ionizing star lies on the
line of symmetry. however, the position of the star at 1.90$\mu$m\, 
shows that the source lies  closer to D2 than D1.

\subsection{Sgr~A East HII A-C}

Unlike the compact HII source D, sources A, B, and C appear rather similar,
both morphologically, as seen in 
Figures 2a,b, and kinematically, in the [Ne II] data cube.

Morphologically, source A appears to have  cometary 
structure  similar to that of G29.96-0.02 \citep[e.g.][]{fey95}.
To test this suggestion, position-velocity diagrams of RA cuts through
source A (as well as B and probably C). 
Figure 8a  shows the position-velocity  diagrams for sources A,  B and C.  
Sources B and C appear to have smaller extents than source A both spatially and kinematically.
Sources A and B appear to have  cometary structure whereas
source C is not resolved. 
The position-velocity diagrams of RA cuts through HII source A
at five different declinations are shown in Figure 8b.
The cuts along the northern and southern rims of the shell are
narrower spectrally than the cuts passing through the center of the HII region,
as shown in middle panels.
The broad lines in the eastern rim of the shell and double lines
near the center of the shell show a pattern that is very similar to that seen 
in many UC HII regions \citep[e.g.][]{zhu08}.
The characteristic position-velocity diagrams of several types 
of compact HII regions shown in Figs. 45-48 of \citet{zhu08},
show that the kinematics of source A is clearly cometary. 
Position-velocity
diagrams of all three sources, with 3\arcsec\ wide cuts, are shown 
in Figure 9.  The position velocity (PV)  diagram for source B is similar to that for A,
but shows a smaller extent in both position and velocity, so is not
so well resolved.  Source C may be similar, but is even less well
resolved.

The PV diagrams of Figure 8 match bow shocks in which the stellar wind of an O star sweeps the 
ionized gas into a thin, paraboloidal shell. The position-velocity diagrams of sources B and C are 
very similar to that of source A. Most compact HII regions with recognizable kinematics observed 
in [Ne II] fall into the latter category, with the HII regions lying on the near sides of their 
molecular clouds, presumably because extinction through molecular clouds prevents observation of 
[Ne II] emission from those lying on the far sides. HII source A (as well as B and probably C) is 
unusual in that the orientation of its position-velocity diagram indicates that its head is tipped 
toward us.  If it lies on the surface of a molecular cloud with its tail pointing out of the 
cloud, it must lie on the back side.  This is somewhat surprising given the small extinction to 
the source.  In addition, its velocity indicates that the star is moving relative to the 50 \kms\ 
molecular cloud.  In fact, its kinematics matches very well that expected for ionized gas in a bow 
shock around an O star moving through molecular gas \citep[see][]{wilkin96,zhu08,arthur06}.  The star is moving to the east an toward us at an angle of roughly 30\degree. 
Perhaps this shouldn't be surprising, given that this is the most commonly cited model for a 
cometary HII region, but it is not what was observed by Zhu et al., who found that most cometary 
HII regions are actually pressure-driven flows around stationary stars near the surfaces of 
molecular clouds.  It seems likely that the turbulent environment near the Galactic center with 
high density gradient in molecular gas resulted in stars which rather quickly moved out of their 
natal clouds and are now drifting through lower density (and so lower extinction) molecular gas.


The ionizing stars for A, B, and C are not 
evident in IRAC data (see Fig. 4) . If present, they are not noticeably brighter
than typical fore/background stars, nor are they  
located symmetrically at the centers of A-C.
In stark contrast, D has an extremely bright point source.
This difference alone indicates a significant difference between
sources D and A-C.
Additional difference between D and A-C HII regions comes from  
MSX observations. Source D has 
the extremely bright point source that is increasingly more prominent with increasing wavelength. 
However, the relative brightness of D drops noticeably 
compared to A-C when  MSX data at 8 and 24$\mu$m are used. 
This suggests a lower dust mass for D, or that the bulk of the dust is warmer. 

\section{Discussion}

\subsection{The Source of Ionization in Source D}

The radio measurements indicate that D1 is optically thin at 5-44 GHz,   
although the slight increase in flux from D2  at higher frequencies
suggests that it may be partially optically thick at the very core.
The 1.9\um\ measurements also indicate clearly a bright stellar source   
near the peak of the radio emission of D2.
As the emission is predominantly optically thin in radio wavelengths, 
so adopting a
distance 8\,kpc and an ionized gas temperature of 8000\,K, the total
flux of 81\,mJy at 43.3\,GHz yields an estimate of the volume emission
measure $\int n_e^2 \, dV \approx 2\ee 60 \,$cm$^{-3}$.  The associated
hydrogen recombination rate requires an ionization rate $N_\mathrm{L}
= \alpha^{(2)} \int n_e^2 \, dV \approx 6\ee 47 \,$s$^{-1}$, where
$\alpha^{(2)}$ is the hydrogen recombination coefficient into all levels
but the ground state.  As some ionizing photons might escape the
immediate surroundings, be absorbed by dust or be absorbed very close
to the exciting star, where free-free optical depth may be large, 
this estimate is a lower limit on the ionizing photon production rate
by the central source. 
The parameters of the SED of this source suggest that the stellar source 
G359.23-0.063 is associated with a 25 \msol\ star (see Table 3) and 
8$\times10^4$ \lsol.   These parameters  compare well with the $\approx
(0.8-1.5)\times 10^{48}$\,s$^{-1}$ emitted by  a O9/B0.5 star  
\citep{vacca96}. 
The full extent of 
the source in the NS direction is about 18$''$. Assuming a source size of  9\asec and an expansion 
velocity of $\sim 30$ \kms\ give an age of $\sim10^4$ years (This age estimate 
could be  a lower limit 
if the motion  is due to flows along the surfaces of the cloud.)  
This 
presents the earliest phase of massive star formation in the 50 \kms
cloud.

The estimate for the mass of stars exciting sources A-C can not be made 
as the stellar source is not identified in these HII regions. However, if we 
assume that the A-C sources have similar  masses to that of D, we can make an estimate 
of the total mass of stars in this regions.
To compute the total cluster mass of Sgr A East, we adopt 
a standard broken
power-law form of the IMF \citep{kroupa01} and assume  
100  \msol\  in stars above 20 \msol\,. This corresponds to 
a total stellar mass above 0.5 \msol\ of 940 \msol\ associated with 
a group of young stars in Sgr A East. 
Assuming that the typical age of the cluster is  $\sim10^5-10^6$ 
years, we
estimate star formation rate of 0.01 to 0.001  \msol yr$^{-1}$. We can also 
make an estimate of the total mass of molecular gas that went into 
forming stars. We used 850 $\mu$m images of Figure 1 
to estimate the depleted gas mass  by converting submillimeter flux to 
the  mass of molecular gas.  We estimate a   subtracted flux of 15 Jy at 850$\mu$m 
depleted from the region where the group of HII regions is distributed.
We assumed that dust emission from the 50 \kms\ cloud was uniformly distributed 
 inside and outside the HII regions prior the  formation of stars.  
For a dust temperature of 20 K and a conversion factor from 850$\mu$m to 
gas mass \citep{price00}, we   find a  rough 
total gas mass of 7.7$\times10^3$ \msol\ that dissipated and converted into 
940 \msol\ of stars. 


The near-infrared observations show that despite the ionized gas morphology
suggesting a pair of compact HII region, it is probably a single
source that is exciting both  D1 and D2.
Presumably D1 is produced by
irradiation of a dense clump of gas distributed  within a cavity that has
been partly evacuated of lower density gas by the stellar radiation
field.  In this scenario, the clump has an angular diameter of $\sim
8000$\,AU and lies at a projected distance of $\sim 12000$\,AU from D2,
and so intercepts approximately 10\% of the ionizing photons escaping
the D1 region.  The observed 30\,mJy flux at 43\,GHz requires a total
ionization rate at the clump surface of $\sim 2\ee 47 $\,s$^{-1}$ ,
implying that the exciting source is emitting ionizing photons at a
rate $\sim 2\ee 48 $\,s$^{-1}$.

\subsection{The Kinematics of Source D}

We now attempt to construct a model to explain the observations of
source D.  In particular, we need to explain the spatial distribution
of the ionized gas and infrared continuum emission, as well as the
motions of the ionized gas.

The ionized gas is distributed between
two sources, D1 and D2, each of which is elongated in the NS direction.
It is nearly symmetric about a NS line passing between D1 and D2,
but with the eastern peak, D1, lying somewhat north of the western peak,
D2, with a p.a. $\sim$70\degree.  The infrared continuum, at both 1.9\um\ and
12.8\um\, shows a single peak that lies between the ionized gas peaks,
but probably closer to D2.  The kinematics of the ionized gas can be described
by two probably related structures.  A pair of spectrally broad features
extend to the north from the D1 and D2 peaks.  The eastern feature
extends in velocity from near the molecular cloud velocity redward
by about 30 \kms.  The western feature extends about 30 \kms\ blueward
of the molecular cloud.  Lower velocity emission is seen both north and
south of the peaks, broadening both spectrally and spatially going to
the north.

The two broad-lined emission features seem most naturally explained
by a bipolar jet pair, probably interacting with the wall of a
cavity.  The jet could originate from a young star surrounded by
a disk, lying between D1 and D2.  The disk axis is tipped at a
p.a. $\approx$ 70\degree , with the west side tipped toward us,
so that the western jet is blueshifted.  The fact that the jets
are not seen extending to the east and west of D1 and D2 indicates
that the observed [Ne II] emission is not from the jets themselves,
but from the region of interaction between the jets and a surrounding
wall.  In fact, the bright emission peaks could be simply regions of
a wall which are illuminated by ionizing radiation, but the width of
the emission lines indicates that jets are needed.
The offset between the position of the 1.9 \um\ continuum and the position of the axis of 
the symmetry of D1/D2 
can also be explained by  inhomogeneity of gaseous material in the vicinity of the star and/or the 
motion  of the star.  
Figure 9 shows a schematic diagram of the relative location of 
D1,  D2 and the central massive star with a flaring disk.  
The fact that D1 is redshifted and D2 is blueshifted could be
explained if the disk is seen at an angle between edge-on and
face-on, with the west side of the disk tipped toward us. The lack of symmetry in
this diagram is likely the result of density gradient surrounding the star. 
The 50 \kms molecular cloud lies mainly  to the west of the central star. 

High mass stars efficiently photoevaporate their circumstellar disks 
on
timescales of 10$^4$ -- 10$^5$ years.  The piling up of the resulting disk
wind as it expands against the surrounding medium could be responsible for
the $\sim 0.1$\,\msol\,  of ionized material we identify as sources D1 and D2.
The semianalytic photoevaporated disk-wind model of \citet{hollenbach94} applied to a 28\,M\solar\,  star yields a mass flux $\sim
3\times10^{-5}$\,M\solar\,yr$^{-1}$, indicating an age of $\sim3\times10^4$
years.  The momentum of the material  would be maintained by contributions
from the momentum fluxes $\dot{M}v\sim3\times10^{-4}$\,M\solar\,km\,s$^{-1}$ from 
each of the disk
wind ($v \ga 10$,km\,s$^{-1}$)  \citep{drew98,hoare06} and
stellar wind ($\sim3\times10^{-7}$\,M\solar\,yr$^{-1}$ and  v 
$\sim$ 1000 \kms ).   Another  potential  contribution could arise from a fraction of the total
momentum flux $10^{5}$\,L\solar$/c \sim1\times10^{28}\sim
2\times10^{-3}$M\solar\,yr$^{-1}$\,km\,s$^{-1}$ in the photons emitted by
the central star.

The extensions to the north
could trace the regions of interaction of the jets with the cavity
wall if the star is moving to the south, although it is not apparent
why broad-lined emission would persist after the jets pass by.
An explanation is also needed for the probable offset of the infrared
continuum emission toward D2.  One possibility is that the disk that
collimates the jets is close enough to edge-on that it prevents a
direct view of the central star even in the infrared.  The near
infrared continuum radiation could be scattered into our line of sight
by the material that forms the ionized gas peak D2.
Since D1 lies on the far side of the disk, backscattering (which
would be less efficient) would be required for it D1 to be seen in
scattered light.  It may also be obscured by the disk if the disk
extends out far enough to cover it.  The 12.8\um\ continuum is
unlikely to be scattered, but it could also be affected by
extinction, and it is not so convincingly offset from the center of D
as is the 1.9\um\ emission.

The spectrally narrow emission extending both north and south of the D peaks
is even more difficult to explain than the broad-lined emission.  The  
ring-like pattern seen in the position-velocity diagrams in Figure 7 could
be due to gas in each cut lying on an expanding ring. The fact that the PV
rings increase in spatial and velocity extent going to the north would
indicate that the three-dimensional ionized gas distribution is on the
surface of an expanding cone, with the expansion speed increasing to the
north. Alternatively, the gas could be flowing along the surface of a cone
if it is accelerating so as to make its speed increase linearly with
declination. However, we cannot propose a physical model that would cause
either of these flow patterns. We know of no other case of an expanding cone
like what we suggest or any reason a conical wall would expand in this way.
A flow along the surface of a cone seems more natural, but all cometary HII
regions we have observed, with sources A-C being representative, are
paraboloidal rather than conical, with much larger opening angles that in D.
They also have most of their acceleration in a small distance from their
vertexes. (A constant $dv/dt$ would result in $v^2 \propto r$, or $dv/dr
\propto r^{-1/2}$, rather than the observed constant $dv/dr$.) It is
especially puzzling that the ionized cone has its vertex 3\arcsec\ south of
D, so presumably leading the ionizing star, which we think is moving in that
direction. We have to conclude that we are unable to propose a consistent
model for the north-south  velocity of ionized gas associated with source D. The
narrow-lined gas appears to be distributed on the surface of a wedge or cone
with an opening angle $\sim$20\degree , with its vertex $\sim$3\arcsec\
south of the star.

\subsection{The Kinematics of Source A}

Why are cometary HII sources A-C oriented head-on, with velocity
offsets indicating that their ionizing stars are moving through
the 50 \kms\ molecular cloud, whereas most cometary HII regions
seen in [Ne II] are seen tail-on with little stellar motion?
The answer may lie in the circumstances of their formation.
The location of the Sgr~A East HII regions near the edge of
the Sgr~A East supernova remnant suggests that the ionizing
stars may have formed as a result of compression of molecular gas
by the SNR.
If the stars formed in swept-up gas, they would have formed with
the velocity of the gas.  After their formation, the compressed shell
would have slowed as more gas was swept up, and the stars could have
drifted out of the compressed gas.
The stars may now be moving through gas of low enough density that
extinction does not affect the [Ne II] emission, as it would if
they were still inside of a dense molecular cloud.
From the orientation of its broken shell, it appears that the ionizing
star of HII source A is moving to the east, as well as toward us.
From the offset of its position-velocity diagram, its motion relative to the
50 \kms\ molecular cloud is $\sim$30 \kms.
The ionizing stars of sources B and C have similar motions.
In particular, the position-velocity diagrams indicate that all three
sources are moving toward us, while source B is moving to the east and
source C is moving to the northeast.

The $\sim 50$\,km\,s$^{-1}$ velocity width of the [Ne\textsc{ii}] line
is consistent with other ultra-compact H\textsc{ii} regions 
\citep[e.g.][]{garay99}. 
The characterization of a cluster 
of HII regions Sgr~A A-D may not be correct, as these HII regions appear 
to be unbound due to their large physical separation,
signifying isolated star formation near the Galactic center. 
Again, the site of on-going star formation is  consistent 
with  sites of massive star formation elsewhere \citep[e.g.][]{garay99}. 

\subsection{Conclusions}

We have presented multi-wavelength observations of a chain of HII regions located 
at the eastern edge of the Sgr A East supernova remnant  and the 50 \kms\ 
molecular cloud. These HII regions  show the youngest   star formation activity closest to the 
Galactic center.   
The  youngest member  of Sgr A East HII  regions 
G359.956-0.08 or the Sgr A D source is estimated to have an age of roughly 
1.5$\times10^4$ years and consists of two elongated ionized features. 
We  identified   the  central star 
responsible for  ionizing 
both components of  the D source. The SED fit of the central star indicates
a mass of 25 \msol\  and a luminosity of 8$\times10^4$  \lsol\ . 
The kinematics of ionized gas show a 
an  E-W red and blue shifted flow with respect to the central star and a complex  
asymmetric  velocity structure in the N-S direction.
We presented  a simple  model in which the central star 
is surrounded by a disk constraining the flow of ionized wind.  
In this picture, the UV radiation from the central hot star may generate an 
outflow from the disk, sweeping up the ISM gas and forming the two components of 
ionized gas D1 and D2.  Given that high density molecular material 
is distributed  to the west, the ionized feature D2 is expected to lie closer to the 
central star. Future high spectral resolution 
observations of 
ionized gas and  
detailed modeling of the complex N-S velocity structure  should 
explain better the  kinematics  of ionized gas presented here.

\acknowledgments
This work is partially supported by grants AST-0807400 
(to FYZ) and AST-0607312 (to JHL) from the National Science Foundation. 
This research has made use of the NASA/ IPAC Infrared Science Archive, which is
operated by the Jet Propulsion Laboratory, California Institute of Technology,
under contract with the National Aeronautics and Space Administration.

\vfill\eject

\begin{figure}
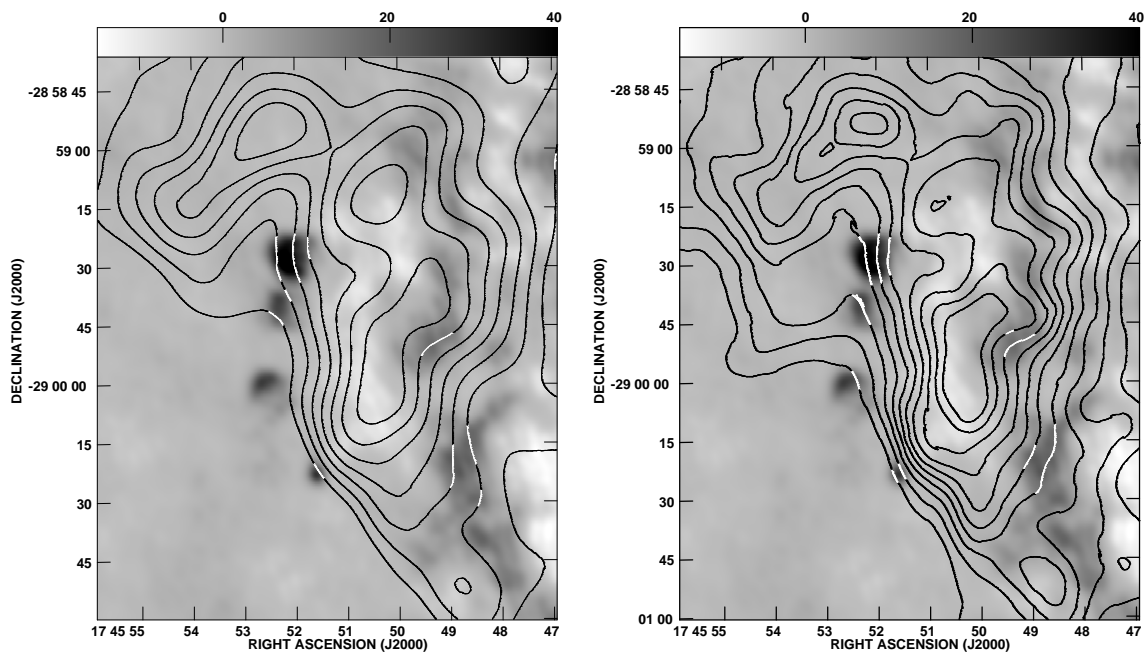

\center
\includegraphics[scale=0.4,angle=0]{f1a_hii-a-d.ps}
\includegraphics[scale=0.4,angle=0]{f1b_hii-a-d.ps}
\caption{
{\bf (a) Left} 
Contours of 450$\mu$m emission from the Sgr A East HII region 
are superimposed on a 20cm continuum image  (black) and are 
represented at levels  32, 34,..., 50 Jy 
beam$^{-1}$ (beam size is 8$''\times8''$). 
{\bf (b) Right } Similar to (a) except that contours of 850$\mu$m emission 
are presented at  levels (32, 34,.., 46 )$\times 0.2$ Jy beam$^{-1}$ (beam size is 15$''\times15''$). 
The greyscale 20cm continuum image (in black) has a resolution of 
3.5\arcs\ and 2.9\arcs\ (PA=26$^\circ$). 
The 450$\mu$m  and 850$\mu$m emission trace  dust
emission from the 50 \kms\,  molecular cloud GMC M-0.02-0.07. 
}
\end{figure}

\begin{figure}
\center
\includegraphics[scale=0.35,angle=0]{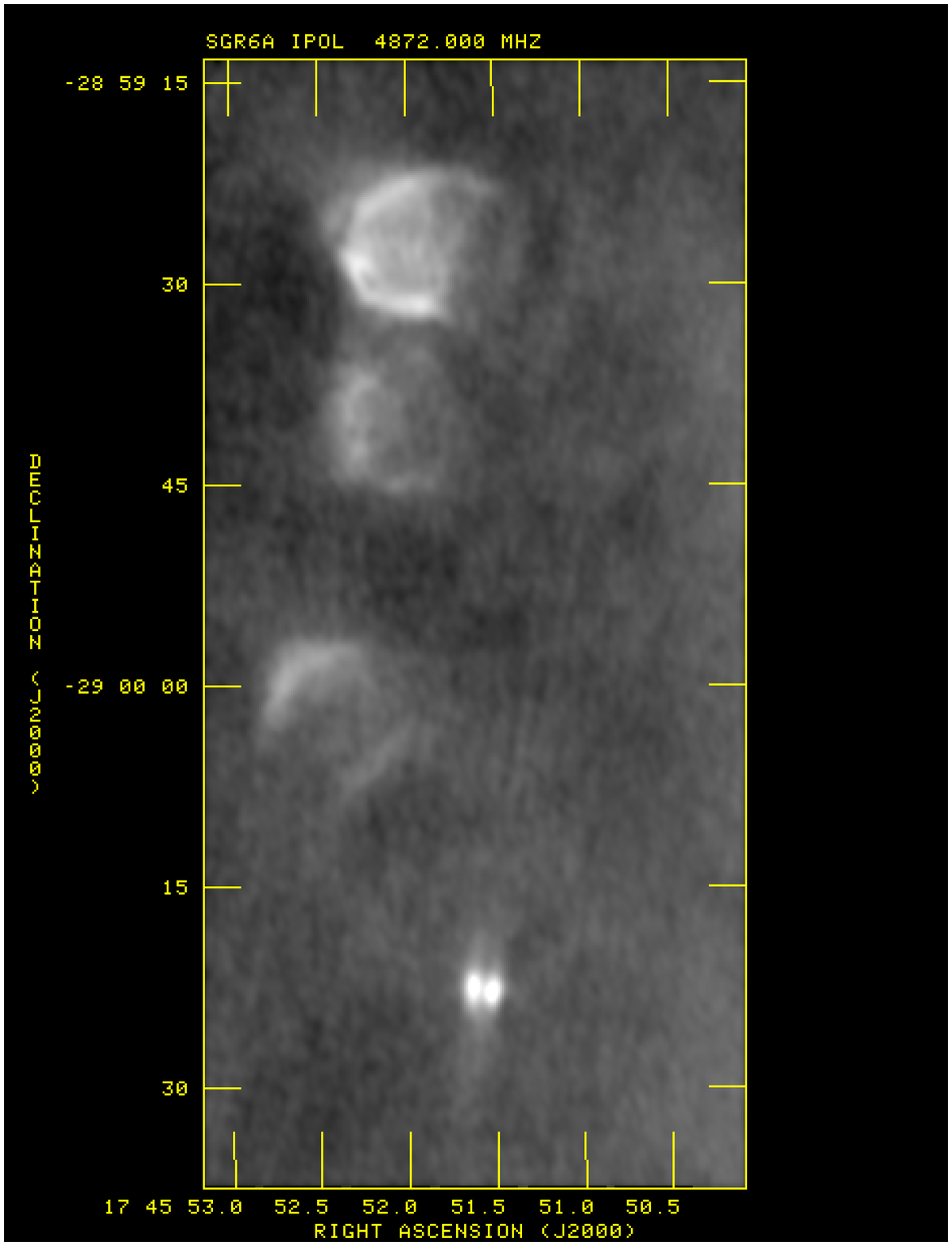}
\includegraphics[scale=0.35,angle=0]{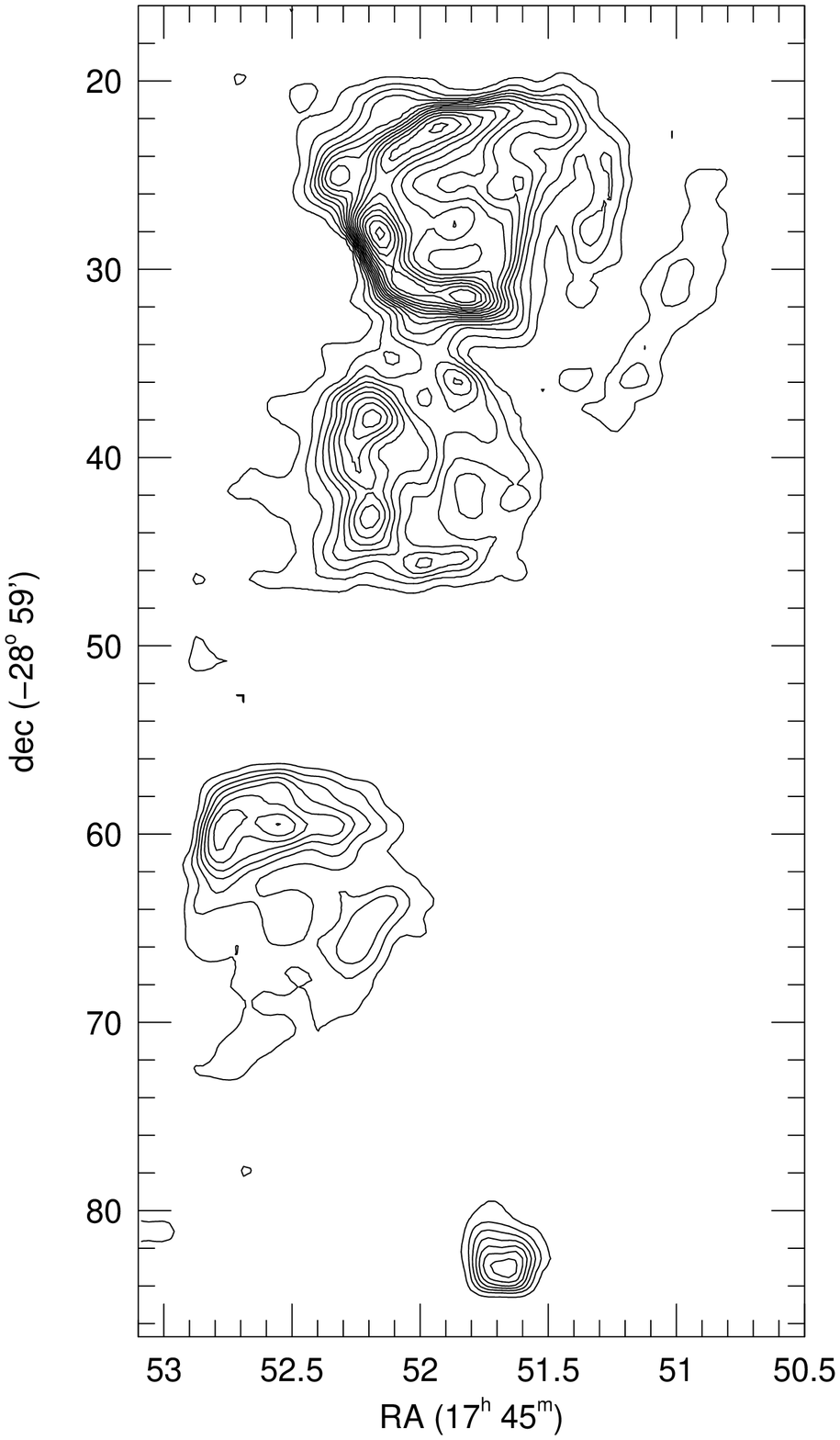}\\
\clearpage
\includegraphics[scale=0.5,angle=0]{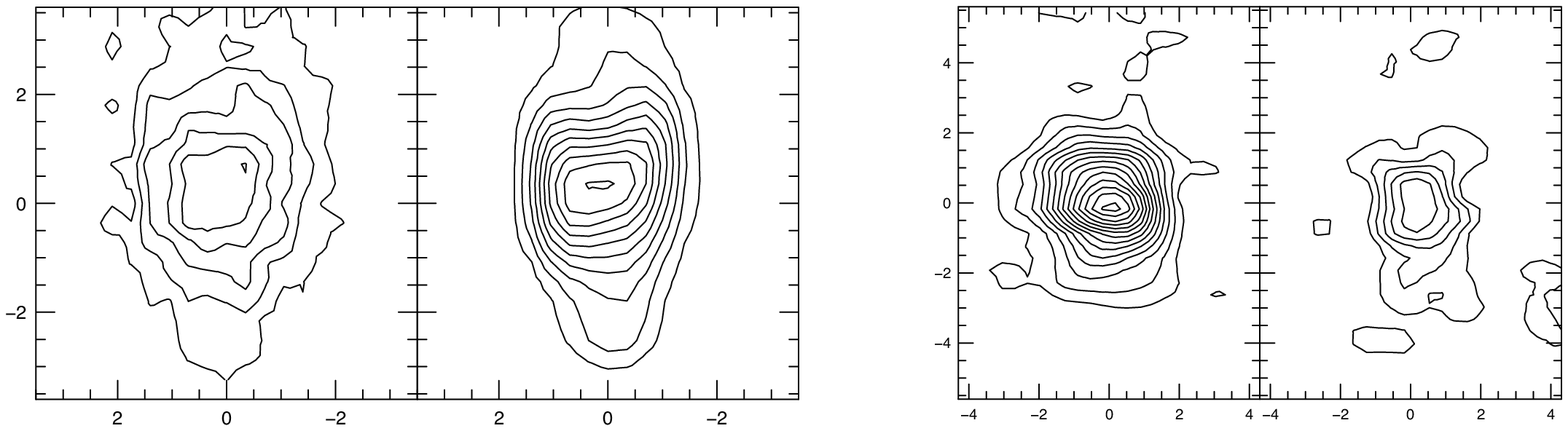}
\caption{{\small
{\bf (a) Top Left} 
A 6cm continuum image of the Sgr~A East HII region with a spatially
resolution of 0.83\arcs\ and 0.52\arcs\ (PA=11$^\circ$).  The top three
HII sources A to C are resolved having shell-like structures whereas
the most compact HII source D shows two compact sources elongated in
the north-south direction.
{\bf (b) Top Right} 
The same as (a) except that the map is based on
integrated [NeII]  line emission over a velocity range of 
V$_{LSR}$ = 10 to 70 \kms\ corresponding to spectral interval of 0.158 cm$^{-1}$.  
Contours of the line emission are  set 
quadratically at (9, 16, 25, ...) $\times 5\times 10^{-4}$  
erg s$^{-1}$  cm$^{-2}$  sr$^{-1}$.
{\bf (c) Bottom Left} 
Contour map of source D  before and after deconvolution 
averaged over a 70 km s$^{-1}$  interval centered at 50 km s$^{-1}$ with contour  
intervals 0.02 erg s$^{-1}$  cm$^{-2}$ sr$^{-1}$.
{\bf (d) Bottom Right}
The line and 
continuum maps are shown in the left and right panels 
with contour intervals of  
2.6$\times10^{-3}$ and 3$\times10^{-3}$
erg s$^{-1}$  cm$^{-2}$  sr$^{-1}$, respectively. 
The line map  is the average over a 200 km/s interval centered on the line. 
}}
\end{figure}  

\begin{figure}
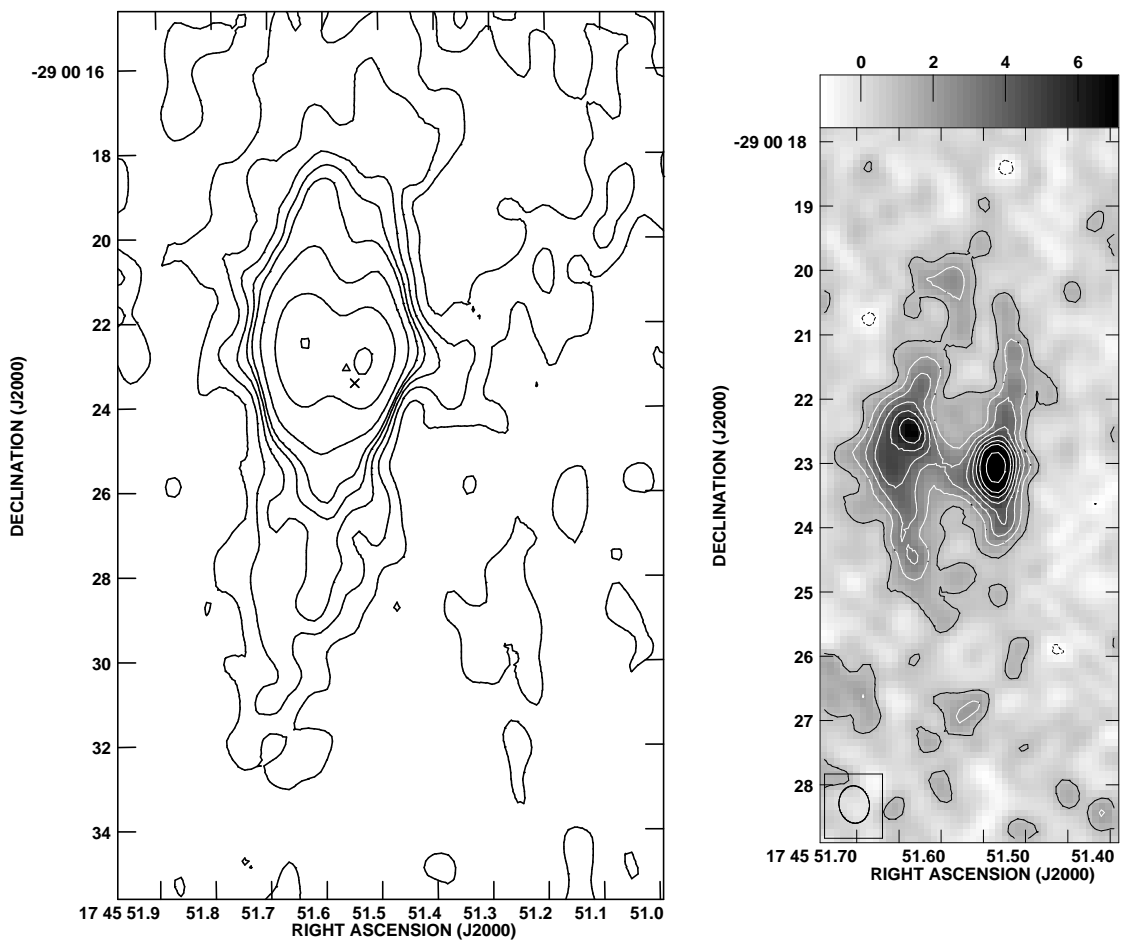

\center
\includegraphics[scale=0.5,angle=0]{f3a_d_hh_6cm.ps}
\includegraphics[scale=0.5,angle=0]{f3b_7mm_d.eps}
\caption{
{\bf (a) Left} 
Contours of 6cm emission from source D are shown with
levels set at (1.25, 1.5, 1.75, 2, 2.25,  3, 6, 15) $\times
10^{-4}$ mJy per beam. The cross and the triangle  show
 the peak positions of 
the 1.90$\mu$m emission 
  and the catalogued position of IRAC source 
with an uncertainty of 0.1$''$ to $0.2''$, respectively.  
Both sources lie  to the southeast of the peak radio emission from D2. 
{\bf (b) Right} A grayscale image of source D
at 43.3 GHz with a spatial resolution of 0.58\arcs\ and 0.47\arcs\
(PA=8$^\circ$).  Contours are set at -1, 1, 2, 3, 4, 5, 6, 7, 8, 9 mJy per
beam.  }
\end{figure}  

\begin{figure}
\center 
\includegraphics[scale=1.5,angle=0]{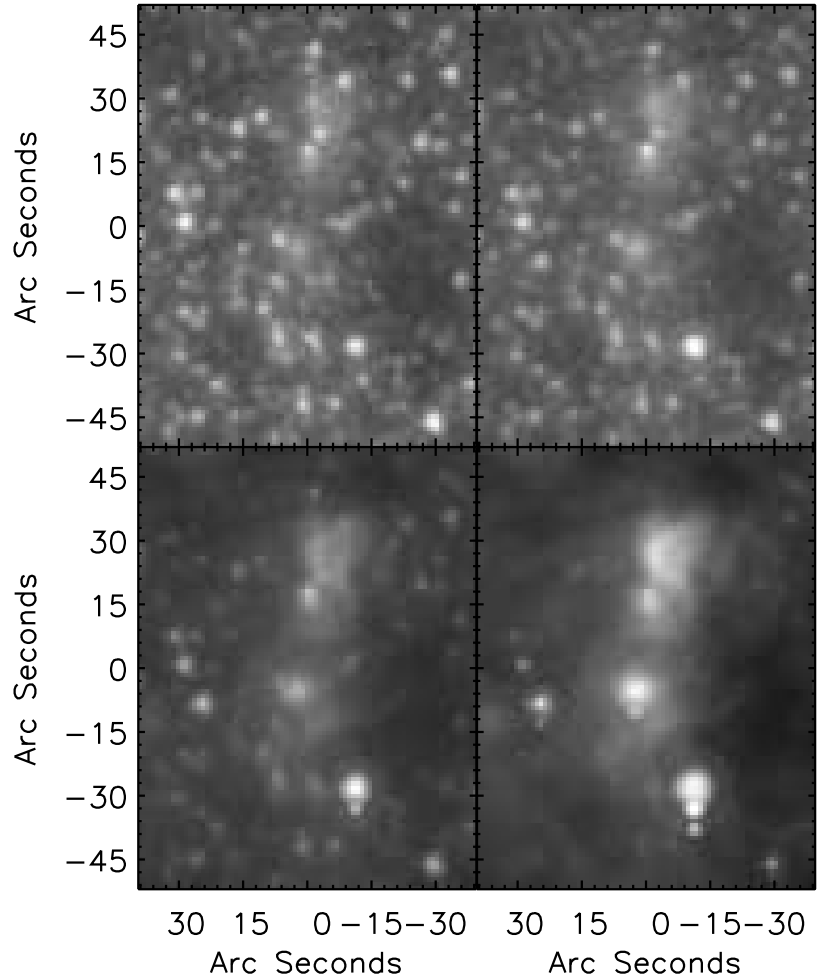}
\caption{{\it Spitzer} IRAC images of the Sgr A East H II regions A--D.
These logarithmically scaled images show emission 
at 3.6, 4.5, 5.8, and 8 $\micron$ (left to right, top to bottom).
The pixel size is $1''$. Bright sources at 5.8 and 8 $\micron$
exhibit detector artifacts $5''$ and $10''$ to their south.
This is especially distinct for Sgr A East D, which is the brightest
source in the field and is obviously saturated at 8 $\micron$.
}
\end{figure}  

\begin{figure}
\center 
\includegraphics[scale=0.6,angle=0]{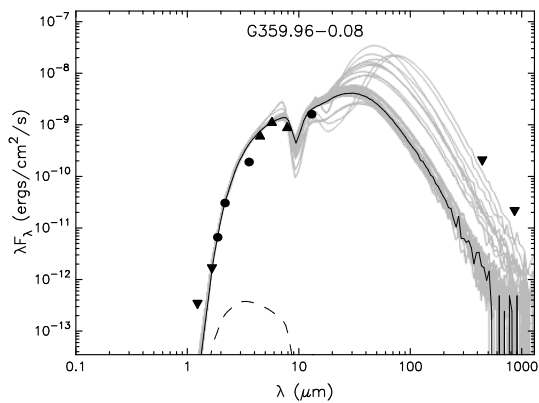}
\caption{ SED fits to source D.  The closed circles are data points, 
downward facing triangle upper limit, upward facing triangles lower 
limits.  The grey 
lines are all models that fit
with $\chi^2 < 3$  higher than the best fit model.  The grey 
dashed  line is the input stellar
spectrum extincted by the best fit foreground extinction.
}
\end{figure}


\begin{figure}
\center 
\includegraphics[scale=0.5,angle=90]{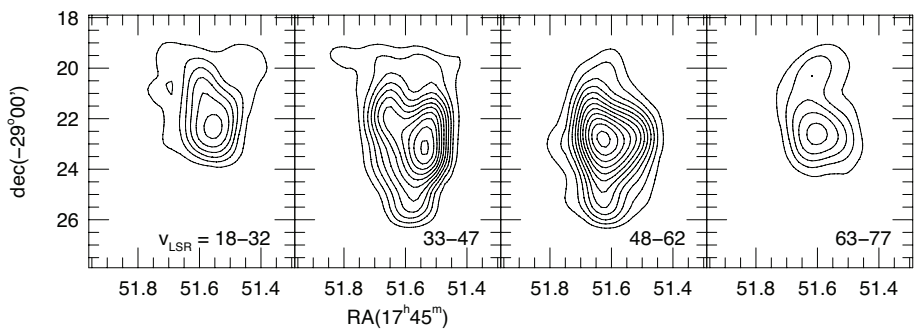}
\caption{Four [NeII] channel maps showing the 
distribution of  ionized gas  with contour
levels set quadratically at (9, 16, 25, ...) $\times 3\times 10^{-3}$  erg s$^{-1}$  cm$^{-2}$  cm  sr$^{-1}$.
}
\end{figure}

\begin{figure}
\center 
\includegraphics[scale=0.6,angle=90]{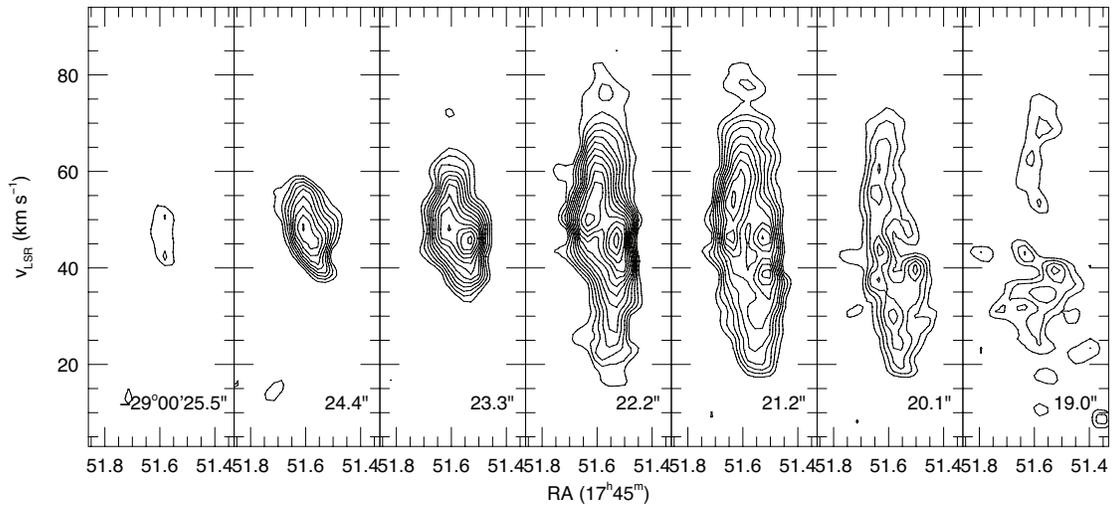}
\caption{
Contours of  right ascension  velocity diagrams  of Sgr~A  D are shown in  seven   different 
declinations. 
The levels  of  integrated [NeII] line intensity 
are set  quadratically at 
(9, 16, 25, ...) $\times 1.9\times 10^{-3}$  erg s$^{-1}$  cm$^{-2}$ cm sr$^{-1}$. The absolute coordinates of 
[NeII] line images  are done by comparing  radio and [NeII] images with each other. 
}
\end{figure}  

\begin{figure}
\center 
\includegraphics[scale=0.7,angle=0]{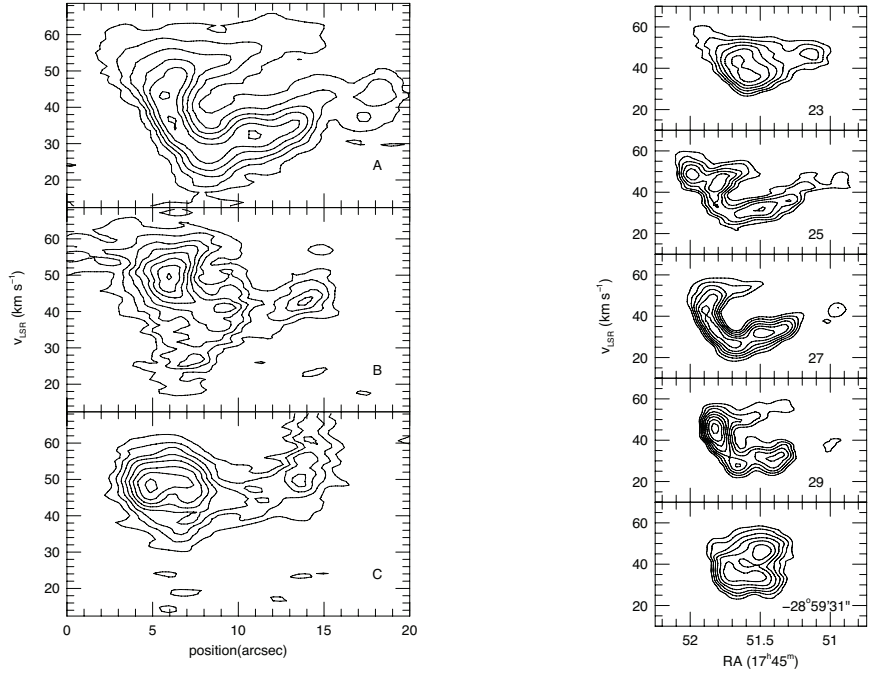}
\caption{
{\bf (a) Left } 
The position-velocity diagrams for sources A,  B and C.  For A and B, the cut runs
EW (E at top of figure).  For C it runs at 45 degrees, NE-SW.  All run near the centers
of the sources, with 5" wide averaging bands.  
{\bf (a) Right} 
Four [NeII] position velocity diagrams of  Sgr~A East A source showing the
velocity cuts at constant declination of (-29$^0 59'$) of 
31.2\arcsec\,   29.1\arcsec\,   26.9\arcsec\,   24.8\arcsec\,   22.6\arcsec\,.
Contour levels are set 
quadratically at (9, 16, 25, ...) $\times 2.7\times 10^{-2}$  erg s$^{-1}$  cm$^{-2}$ cm  
sr$^{-1}$.
}
\end{figure}  

\begin{figure}
\center 
\includegraphics[scale=0.3,angle=0]{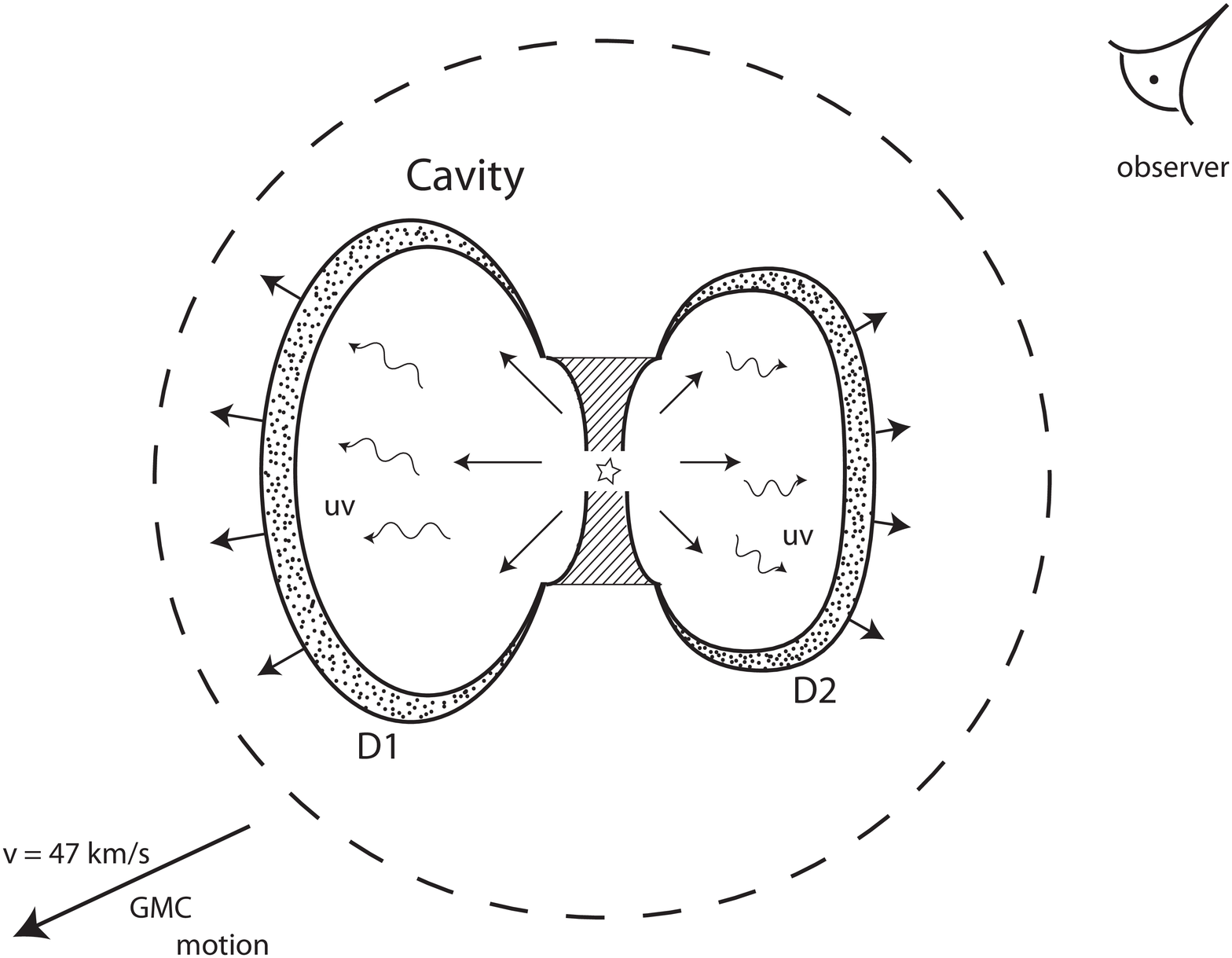}
\caption{A schematic diagram of the features 
of the D source  is drawn. The swept up gas heated
by  UV radiation and
the mass loss from the  central 
young massive star and its disk are likely responsible  for 
producing D1 and D1. The smaller size of D2 is considered to be due to 
inhomogeneity of the material surrounding the source. 
}
\end{figure}

\end{document}